\documentclass[aps,pre,twocolumn,superscriptaddress,groupedaddress]{revtex4}

\usepackage{graphicx}  
\usepackage{amsmath}
\usepackage{bm}        
\usepackage{amssymb}   

\usepackage{float}

\usepackage{color}

\begin{document}

 \title{Removing grain boundaries from three-dimensional colloidal crystals using active dopants}
%
 \author{B. van der Meer}
 
  \author{M. Dijkstra}
   \author{L. Filion}

\affiliation{Soft Condensed Matter, Debye Institute for Nanomaterials Science,Utrecht University, Princetonplein 5, 3584 CC Utrecht, The Netherlands}

\begin{abstract}

Using computer simulations we explore how grain boundaries can be removed from three-dimensional colloidal crystals by doping with a small fraction of active colloids. We show that for sufficient self-propulsion, the system is driven into a crystal-fluid coexistence. In this phase separated regime, the active dopants become mobile and spontaneously gather at the grain boundaries. The resulting surface melting and recrystallization of domains result in the motion of the grain boundaries over time and lead to the formation of a large single crystal. However, when the self-propulsion is too low to cause a phase separation, we observe no significant enhancement of grain growth.  


\end{abstract}
 
\maketitle

\section{Introduction}

One of the key challenges in colloidal self-assembly involves circumventing the long time-scales required for systems to reach their equilibrium states. For example, in colloidal systems where a crystalline phase is thermodynamically stable, the self-assembly of such a phase typically requires crossing a high free-energy barrier. Hence, both nucleation of the equilibrium crystalline phase and the annealing out of crystal defects often occur on prohibitively long time-scales.

An exciting and promising new avenue to aid colloidal systems in crossing free-energy barriers stems from the recent development of active colloidal particles~\cite{paxton2004catalytic,dreyfus2005microscopic,howse2007self,theurkauff2012dynamic,buttinoni2013dynamical,palacci2013living,bechinger2016active}.
 Such particles actively use energy from their environment to generate directed motion and, as such, systems containing active particles are inherently out of equilibrium. Recent experimental and simulation studies have shown that the dynamics of passive systems can be altered dramatically by the incorporation of active particles~\cite{ni2014crystallizing,kummel2015formation,stenhammar2015activity, accepted,wysocki2016traveling}.
Interestingly,   even for very low concentrations ($<1\%$), where the self-propelled particles can be viewed as active ``dopants'', they can help the underlying passive system to equilibrate by facilitating the crossing of free-energy barriers.  Specifically, it has been shown that active dopants can both assist in the nucleation of crystalline domains in high-density hard-sphere glasses~\cite{ni2014crystallizing}, as well as in the removal of grain boundaries in two-dimensional colloidal crystals~\cite{accepted}. 

Here we return to the question of grain boundary removal, and explore whether active dopants can also speed up the removal of grain boundaries in {\em three dimensions}.
In comparison to the 2d case, particles in 3d colloidal crystals are typically more strongly caged,  because the coordination number is higher, and as a result the free-energy barriers involved in annealing out grain boundaries and other defects are expected to be significantly higher. Hence, it is interesting to determine whether active dopants can successfully overcome these higher barriers, and thus assist in coarsening a 3d colloidal polycrystal. 

It should be noted that large, defect-free colloidal crystals have important photonic uses~\cite{colvin2001opals}, and that the growth of such crystals is typically challenging, depending on methods based on, e.g.  epitaxial growth from a template~\cite{van1997template,lin2000entropically,braun2001epitaxial,allard2004colloidal,jensen2013rapid}, the application of external fields~\cite{palberg1995grain,amos2000fabrication,gokhale2012directional,korda2001annealing}, or temperature gradients~\cite{sun2013fabrication,cheng1999controlled}. 
Hence, the possiblilty of actively removing grain boundaries by the addition of active particles shows great promise in future applications.

\section{Method and Model}

\subsection{Model}

We consider a three-dimensional system consisting of $N$ particles that interact via the purely repulsive Weeks-Chandler-Andersen (WCA) potential 
\begin{equation}
  \beta U_{WCA}(r)= \left\{ 
  \begin{array}{ll}
  4\beta \epsilon \left[(\frac{\sigma}{r})^{12}-(\frac{\sigma}{r})^{6}+\frac{1}{4} \right], & r /\sigma \leq 2^{1/6} \\
  0, &  r /\sigma > 2^{1/6} 
  \end{array}
  \right.
\end{equation}
with $\sigma$ the particle diameter, $\beta \epsilon=40$ the energy scale, and $\beta=1/k_BT$, where $k_B$ is the Boltzmann constant and $T$ is the temperature. From the $N$ particles, we identify a subset of particles $N_a$ which we ``activate'' by applying a constant self-propulsion force $f$ along the self-propulsion axis $\hat{\textbf{u}}_i$. We denote the fraction of active particles by $\alpha = N_{a} /N$. 

We simulate the system using overdamped Brownian dynamics. Thus, the equations of motion for particle $i$ read:
\begin{eqnarray}
\dot{{\bf r}}(t) &=& \beta D_0 \left[  - \nabla_i U (t)+  f\hat{{\bf u}}_i(t)\right] + \sqrt{2D_0} {\bm \xi}_i(t)\\
{\dot{\hat {\bf u}}}_i (t) &=& \sqrt{2 D_r} \hat{\bf u}(t)_i \times  {\bm \eta}_i(t),
\end{eqnarray} 
where the translational diffusion coefficient $D_0$ and the rotational diffusion constant $D_r$ are linked via the Stokes-Einstein relation $D_r=3D_{0}/\sigma^2$. 
Additionally, $\bm{\xi}_i(t)$  and $\bm{\eta}_i(t)$ are stochastic noise terms with zero mean and unit variance. Note that for passive particles $f  = 0$. We measure time in units of the short-time diffusion $\tau=\sigma^2/D_0$.

\subsection{Formation of initial polycrystals}\label{sec:polyinit}
We create starting configurations that are polycrystalline by breaking up the simulation box into polyhedral Voronoi cells which we fill with randomly oriented crystal domains~\cite{schiotz1999atomic}. Specifically, we generate a set of randomly placed grain centers, and fill all space closer to a specific center than to any other center with a randomly oriented face-centered cubic crystal domain. In these configurations, it is still possible that two particles are too close to each other. To avoid these ``overlaps'', we remove one of the particles if the center of masses are separated by less than the effective hard-sphere diameter, as extracted from the coexistence densities in Ref. ~\onlinecite{filion2011simulation}.  Finally, the polycrystal is annealed for $\sim 50\tau$ to ensure that the grain boundary structure is relaxed.   

\subsection{Fluid-crystal coexistence simulations}\label{sec:coexsimulations}

We use direct coexistence simulations to determine the fluid-crystal coexistence densities.  More specifically, the coexistence between the crystal and the fluid is studied in an elongated box of  $L_x=L_y \approx L_z/4$ containing $N=16000$ particles. At the start of the simulation, the entire box is filled with a crystal of uniform density. A fraction $\alpha$ of the particles are then labelled as ``active'' particles. To speed up equilibration, the active particles are chosen in a narrow slab in the box \footnote{Note that we have also performed simulations where the active particles are initially distributed randomly throughout the box, but over the course of the simulations they are excluded from the crystal region and migrate to the fluid region. Hence, starting in this manner reduces the time needed to reach the steady state.}. 
During the  simulation, the crystal partially melts near the region rich in active particles. As a result, the system forms a crystal-fluid coexistence, with the two interfaces perpendicular to $\hat{\bm{z}}$. As the fluid phase has a lower density than the overall system, the density of the crystal region increases. As a result, the crystal becomes compressed along the $z$-direction introducing undesirable stresses in the crystal. To allow these stresses to relax, we developed a simple algorithm to reshape the box during equilibration. Specifically, we measure the virial stress tensor in the bulk of the crystal region. Note that in this region, the fraction of active particles is negligible, and thus the virial stress is expected to be a good approximation of the true stress. We then average the stress in the crystal region over $0.2 \tau$, and subsequently make a small adjustment to the  aspect ratio of the box  which reduces the stress, while keeping the total volume constant. We then continue the simulation and repeat this process until the system is in a steady state and the stress in the crystal phase is effectively zero. 
We then further equilibrate for $100\tau$ with a constant box shape, and finally start measuring the coexistence densities. 

\subsection{Cluster algorithm} \label{sec:cluster}
To distinguish different crystalline domains, we use a cluster algorithm based on the local bond-order parameter $q_6$~\cite{wolde1996simulation}. As a first step, we identify the neighbours for each particle, defined as all particles located within a distance $r_c=1.5\sigma$ of the particle in question. We denote the number of neighbours for particle $i$ as $\mathcal{N}(i)$. The bond orientational order parameter of a particle $i$ is then calculated using
\begin{equation}
q_{6,m}(i)= \frac{1}{\mathcal{N}(i)} \sum \limits_{j=1}^{\mathcal{N}(i)} Y_{6,m}(\theta_{ij},\phi_{ij}),
\end{equation}   
where  $Y_{6,m}(\theta , \phi)$ are spherical harmonics, $\theta_{ij}$ and $\phi_{ij}$ denote the polar and azimuthal angles associated with the center-of-mass distance vector ${\bf r}_{ij}= {\bf r}_j- {\bf r}_i$, and $m \in [-6,6]$. We then quantify the degree of correlation in the environment of two particles $i$ and $j$ using 
\begin{equation}
d_6(i,j)=\dfrac{\sum \limits_{m=-6}^{6} q_{6,m}(i) q_{6,m}^*(j)}{\left(\sum \limits_{m=-6}^{6} |q_{6,m}(i)|^2\right)^{1/2} \left(\sum \limits_{m=-6}^{6} |q_{6,m}(j)|^2\right)^{1/2}} .
\end{equation}
Particles within the same domain exhibit $d_6(i,j) \approx 1$ as a result of the translational symmetry within the local crystal lattice. On the other hand, for pairs of particles at the grain boundary $d_6(i,j)<1$, as the translational symmetry is broken due to the different crystallographic orientations of the domains. 

We define the number of crystalline neighbours $\xi(i)$ of a particle  $i$  as the number of neighbours $j$ with $d_6(i,j) >0.95$. Subsequently, we consider a particle as being part of the crystal phase when $\xi(i) \ge 9$. Finally, we consider two particles to be part of the same domain when they are crystalline neighbours.

%
%
 
\begin{figure*} 
\includegraphics[width=\textwidth]{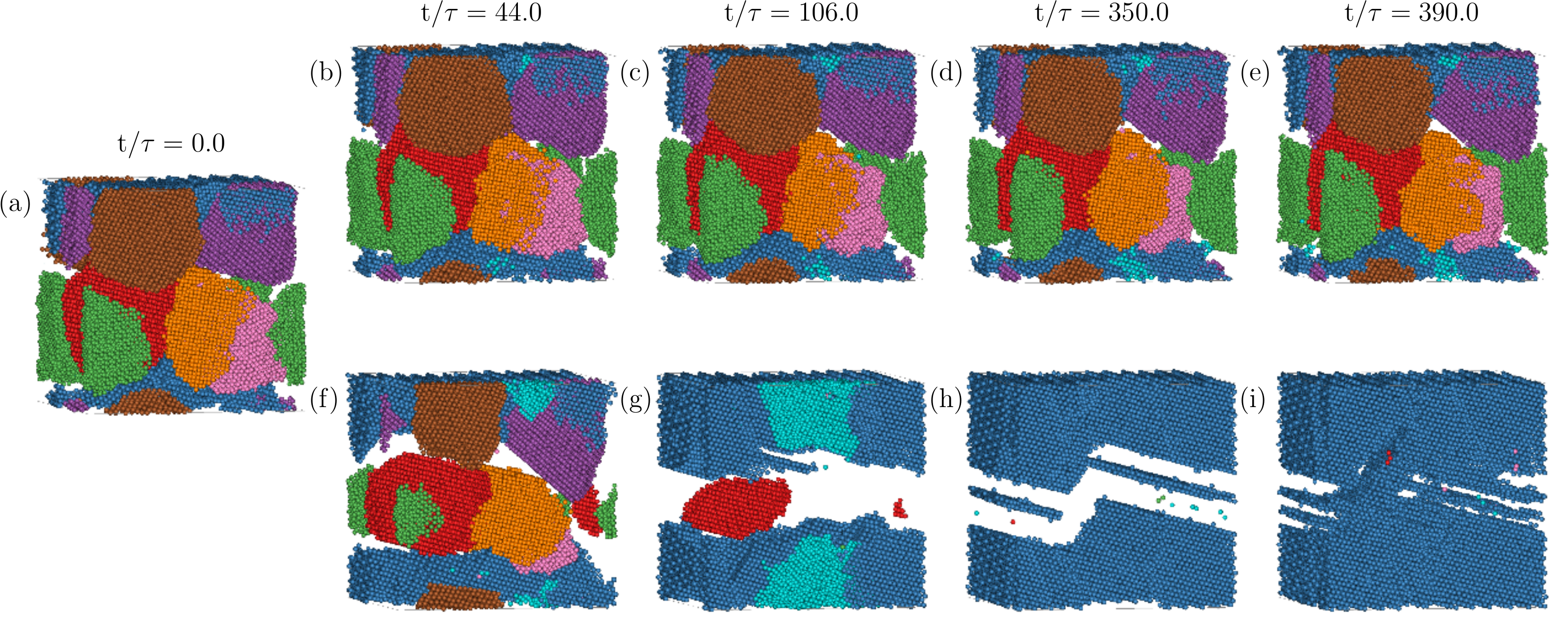} 
\caption{Grain growth in a passive system (a-e) vs a doped system (a,f-i) starting from initial configuration (a), showing significant coarsening of crystal domains in the doped system, compared to its purely passive counterpart. In the lower panel we activate a fraction $\alpha=0.01$ of  particles for $0<t/\tau<350$ with a self-propulsion of $f\sigma /k_BT=90$. After the activity of the self-propelled particles is switched off we can clearly see that the polycrystal has coarsened.  The disordered particles are not depicted.  The system contains  $N=83481$ particles and the  density is  $\rho_s \sigma^3= 0.86$.}
\label{thermalvsdriven}
\end{figure*} 

\section{Results}

In Ref. \onlinecite{accepted}, we showed that it was possible to anneal out grain boundaries in two-dimensional crystals through doping with active particles. Here, we explore whether this approach is also applicable to three-dimensional systems, where the particles are more strongly caged. To this end, we explore the evolution of a polycrystal both in the case of a purely passive system ($\alpha = 0$), and when a small fraction ($\alpha = 0.01$) of the particles is active.  A series of snapshots from both scenarios is shown in Fig. \ref{thermalvsdriven}. Note that in these snapshots, the cluster algorithm described in Section \ref{sec:cluster} was used to identify the crystalline domains in the system, and that particles identified as disordered are not depicted.  As can be seen clearly from the time-series, after $t = 350 \tau$, the passive system still exhibits the same grain structure as in the initial configuration: the polycrystal is trapped in a metastable state and does not evolve to its single-crystal equilibrium state [Fig. \ref{thermalvsdriven}(a-e)]. In contrast, when a small fraction of the particles is active, we observe substantial grain boundary mobility, leading to the rapid coarsening of domains over time [Fig. \ref{thermalvsdriven}(f-h)]. In this case, when $t = 350 \tau$, we are left with a single-domain crystal in coexistence with a fluid.  Thus, similar to the scenario in two dimensions, active dopants help the polycrystal evolve to a large-scale defect-free crystal. Moreover, again similar to the 2d case, upon turning off the activity [Fig. \ref{thermalvsdriven}(i)], the fluid region recrystallizes using the existing crystal as a template. We are left with a single domain crystal with a large stacking fault. Note that this stacking fault is the result of the finite size of our periodic simulation box, as the crystal typically chooses a random orientation incommensurate with the box shape.



\begin{figure*} 
\includegraphics[width=1.0\textwidth]{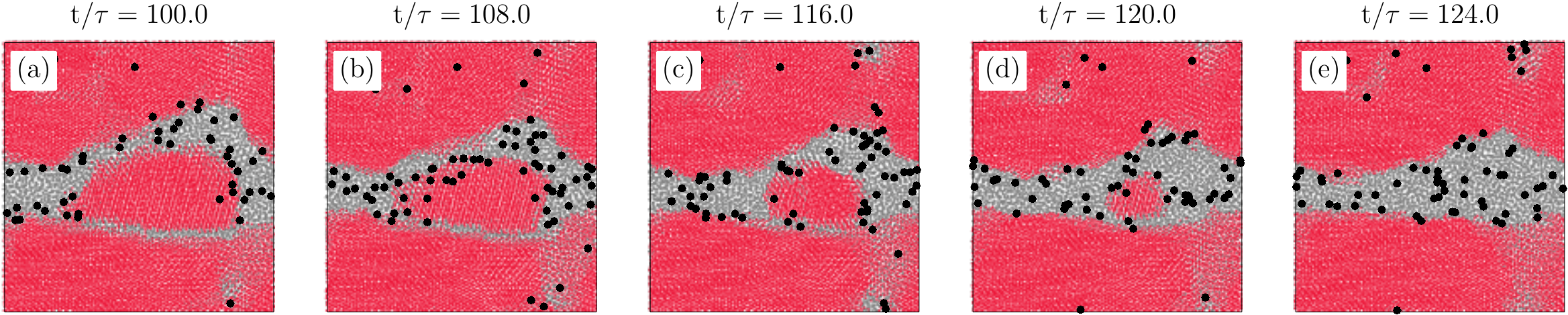} 
\caption{Activity-induced surface melting of polycrystalline domains. We color code disordered and crystalline particles with grey and red, respectively. Active particles are highlighted in black at many times their own diameter. (a)-(e) Over time we can clearly see that the central grain melts from the surface inwards, until it is completely removed. The  active particle fraction is $\alpha=0.01$ with a self-propulsion of $f\sigma /k_BT=90$. The system size is $N=83481$ and the density is  $\rho_s \sigma^3= 0.86$.}
\label{surfacemelting}
\end{figure*}

We now turn our attention to the mechanism responsible for the increased mobility of the crystal grain boundaries. To explore this mechanism, we study the dynamics of the active particles during the coarsening process. In Fig. \ref{surfacemelting} we show the time evolution of a slice of the crystal a short time after the activity is turned on. In this figure, disordered and crystalline particles are coloured grey and red, respectively, and the active particles are highlighted in black. We note that at $t=0$, the active particles are randomly distributed throughout the system. However, it is clear from Fig. \ref{surfacemelting} that by $t = 100\tau$ almost all the active particles are located inside the grain boundaries, which have broadened into thick fluid layers. As time progresses, the domain in the center gradually melts from the outside inwards, while the adjacent domains grow. This continual surface melting and recrystallization of domains is responsible for the motion of the grain boundaries and leads to the formation of the large single crystal in Fig. \ref{thermalvsdriven}(h).

\begin{figure} 
\includegraphics[width=0.50\textwidth]{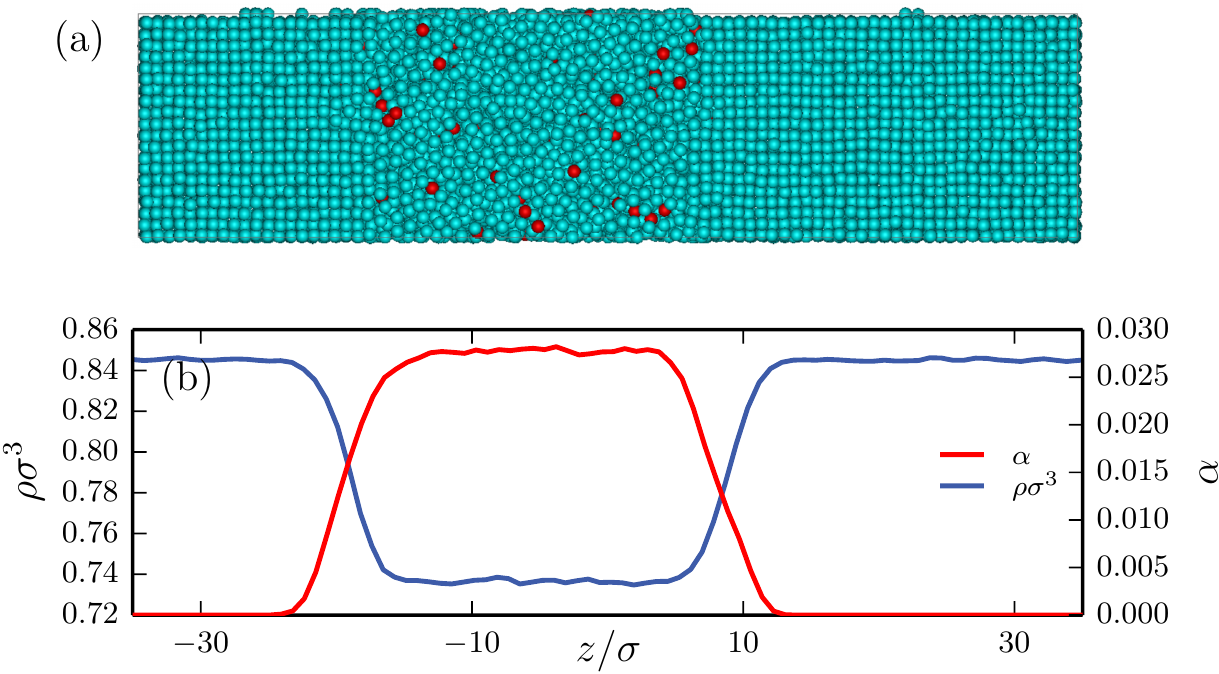} 
\caption{(a) Direct coexistence between a crystal and a fluid phase at overall density $\rho_s \sigma^3=0.80$ and an active particle fraction $\alpha=0.01$ and self-propulsion $f \sigma / k_B T=50$. (b) The corresponding density profile (blue) and the local fraction of active particles (red).}
\label{phasediag}
\end{figure}

The appearance of a substantial fluid region, as in Fig. \ref{thermalvsdriven}(h), suggests that the presence of active particles shifts the phase boundaries in this system. To confirm this, we studied the coexistence between the crystal and fluid using simulations of $N=16000$ particles in an elongated box as discussed in Section \ref{sec:coexsimulations}. 
In Fig. \ref{phasediag}(a), we show an example snapshot, where we highlight active particles in red.
Clearly, the system depicted in this snapshot has separated into a crystal containing (almost) no active particles and a fluid rich in active particles. To analyse this coexistence in detail we measure the density profile $\rho(z)$ and active particle fraction $\alpha(z)$ along the long axis of the box. Specifically, we divide our system into slabs of length $\mathrm{d}z=a$ in the long direction of the box, with $a$ the lattice spacing of the crystal. To calculate the active particle fraction and density, we then determine the mean Voronoi volume per particle in each slab. The resulting profiles, corresponding to the system depicted in Fig. \ref{phasediag}(a), are shown in Fig. \ref{phasediag}(b). Here, two distinct densities are visible, namely a high-density crystal phase largely free of active particles and a low-density fluid phase rich in active particles. One might further ask whether the active particles accumulate at the fluid-solid interface, as they are known to accumulate at walls~\cite{elgeti2013wall,yang2014aggregation}. However, the smooth decay in active particle fraction  (red line in Fig. \ref{phasediag}(b)) between the high and low density phases clearly indicates that the active particles do not specifically adsorb to the crystal-fluid interface. 

\begin{figure} 
\includegraphics[width=0.35\textwidth]{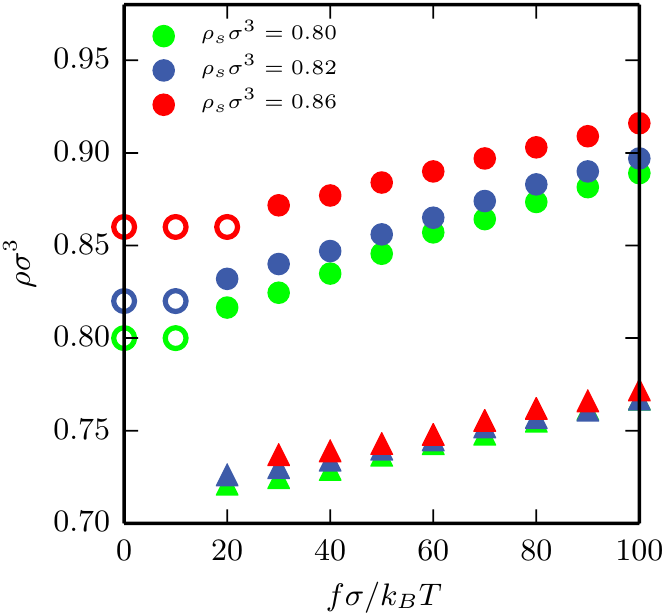} 
\caption{Coexistence densities as a function of self-propulsion $f \sigma / k_B T$,  with a fixed active particle fraction $\alpha = 0.01$ for three different overall system densities $\rho_s \sigma^3$. Open circles indicate the absence of phase coexistence (pure crystal phase). Closed circles indicate the densities of the coexisting crystal phase, and triangles indicate the coexisting fluid densities. System size: $N=16000$.}
\label{phasediag2}
\end{figure}

Density profiles, like the one shown in Fig. \ref{phasediag}(b) can be used to map out the coexistence regions in our system. The results, for a range of overall system densities  $\rho_s$, are summarized in Fig. \ref{phasediag2}. From this figure, we first note that for different overall densities the coexistence curves do not collapse in the $\rho \sigma^3$ vs $f \sigma / k_B T$ representation when the fraction of active particles is kept fixed at $\alpha=0.01$. This feature arises from the fact that our system is in fact a binary mixture and hence should not collapse in this representation. We also see from Fig. \ref{phasediag2} that for low activity, the system does not phase separate for any of the densities we examined.  This was to be expected, since we always chose the overall system density to be higher than the predicted crystal coexistence density for the passive system ($\rho_c\sigma^3 = 0.785$ \cite{filion2011simulation}) as we were interested in removing grain boundaries from crystals.  Importantly, we also see that both the crystal and fluid coexistence densities increase monotonically as a function of the activity of the dopants. Hence, when the activity is sufficiently strong, the melting density is increased beyond the overall system density, and as a result the system partially melts into a dopant-rich fluid and a nearly undoped crystal. We identify this partial melting as the driving mechanism behind the broadening of the grain boundaries observed in e.g. Fig. \ref{surfacemelting}(a). Subsequently, these fluid regions, which display high diffusion rates due to the relatively high concentration of active dopants, enable the fast annealing of the grain boundaries.

%
%



\begin{figure}[t] 
\includegraphics[width=0.3\textwidth]{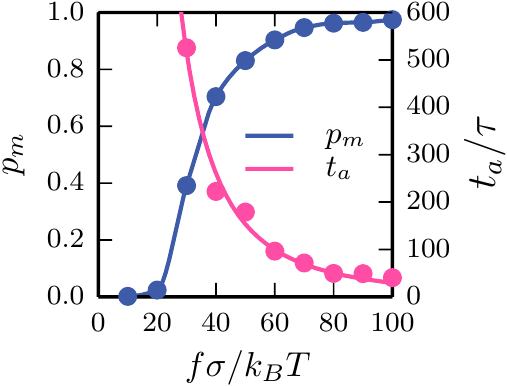} 
\caption{Time scale associated with formation of a large domain $t_a$ and the fraction of mobile active particles $p_m$ as a function of the self-propulsion $f$ for a density $\rho_s \sigma^3=0.86$. }
\label{mob}
\end{figure}

Hence, we identify phase separation into a passive crystal and a fluid rich in active particles as the key mechanism behind active grain growth. To further illustrate this we  will now return to the study of polycrystalline systems and  follow the evolution of grain sizes over time for  polycrystals  of overall system density $\rho_s \sigma^3 = 0.86 $ upon the activation of a fraction of $\alpha=0.01$ of the particles. We now define a time scale $t_a$ associated with the growth of a large domain by calculating the mean time required to form a crystal domain comprising at least a fraction $X_c^*=0.4$ of all the crystalline particles. This cutoff ensures that the domain has formed at the expense of at least three neighbouring domains. Note that we have checked that the results are robust to variation in $X_c^*$. We plot the time scale of annealing $t_a$ in Fig. \ref{mob}. At low self-propulsion $f\sigma/k_BT\lesssim30 $, in the region where no phase separation occurs according to Fig. \ref{phasediag2}, we observe no significant grain growth and thus $t_a$ is far beyond our simulation time. Only at sufficiently high self-propulsion $f\sigma/k_BT\geq30 $, where the system phase-separates, do we observe a huge increase in grain growth, leading to lower values of $t_a$. In this regime, we find that increasing the self-propulsion leads to faster annealing of polycrystalline domains. Depending on the self-propulsion $f$, we also observe that a large fraction of active particles remains kinetically arrested within the crystal phase. 

To quantify the extent of dynamic arrest we calculate the fraction of mobile active particles $p_m$, defined as particles that have moved more than one lattice spacing $\Delta r \ge a$ within a time interval $\Delta t=2.5\tau$  \footnote{We have checked that our results are robust to variations in both the displacement threshold as well as the time-interval.}. Clearly, at self-propulsions $f\sigma/k_BT<30$ where we expect no phase separation, the fraction of mobile active particles $p_m \approx0$: all the active particles remain caged inside the domains. At the minimal amount of self-propulsion required for phase separation,  $f\sigma/k_BT \approx 30 $, we observe a large increase in $p_m$, as the active particles manage to escape the crystal and create a fluid. However, near this increase only a fraction of active particles become mobile and participate in the active grain growth process. With increasing self-propulsion the fraction of active particles that partake in the process increases until it approaches unity, thus leading to lower values of $t_a$.

\section{Conclusions}
 
In conclusion, we have explored how grain boundaries can be removed from three-dimensional colloidal crystals by doping with a small fraction of active colloids. We show that at sufficient self-propulsion, the system is driven into a crystal-fluid coexistence. In this case, active dopants become mobile and spontaneously gather at the grain boundaries. The resulting surface melting and recrystallization of domains is responsible for the motion of the grain boundaries and leads to the formation of a large single crystal. However, if the self-propulsion is too low to cause a phase separation, we observe no significant enhancement of grain growth.

Our study thus demonstrates that, similar to the 2d case~\cite{accepted}, doping with active particles provides an elegant new route to removing grain boundaries in 3d colloidal polycrystals. In both cases, we observe that the active particles are attracted to the grain boundaries. This implies that the active particle autonomously find the location inside the sample where their active properties are most helpful. Importantly, we show that after a single large domain has formed, the activity should be switched off to   recrystallize the fluid regions, using the existing crystal as a template. Note that experimentally one may switch off the activity by using active particles that are responsive to certain stimuli (e.g. light-activated colloids) or simply by letting all the fuel be consumed.

Our results, combined with the recent observation of active particles aiding crystal nucleation in hard-sphere glasses~\cite{ni2014crystallizing},  raise the question whether active dopants can be used in other, hard to equilibrate passive systems. For instance, it will be interesting to discover whether active dopants can help nucleate long sought-after crystals, such as a colloidal diamond phase~\cite{romano2011crystallization}, or the laves phases~\cite{hynninen2007self}.  
 
 \section{Acknowledgements}
We acknowledge funding from the Dutch Sector Plan Physics and Chemistry, and funding from a NWO-Veni grant. We thank Vasileios Prymidis, Frank Smallenburg, Simone Dussi, and Siddharth Paliwal for many useful discussions and carefully reading our manuscript.

\bibliography{bibfile}

\end{document}